# “ChatGPT, help me draft a breakup text”: The Covert Triad and Articulation Labor in AI-Assisted Romantic Communication

Skyler Wang[1], Isabella Luppi
McGill University

## ABSTRACT

Generative artificial intelligence (AI) has begun infiltrating the most ordinary domains of romantic life—drafting apologies, softening reproaches, and decoding a partner’s ambiguous messages. While recent scholarship on AI in intimate life has concentrated on chatbot companions, this article shifts the frame to AI as *an intermediary in human-to-human romantic communication*. Drawing on a multimodal corpus of public accounts, commentary, and representations from 2023 to 2026, we contribute two complementary concepts. The *covert triad* names a structural reconfiguration: a relationship that appears dyadic but operates triadically, with AI visible only to the partner who deploys it. *Articulation labor* identifies the mechanism whereby the expressive component of emotional labor—converting felt experience into language a partner can receive—is delegated to AI, even as feeling labor remains lodged in the user. Authenticity thus moves from linguistic authorship toward emotional ownership, a shift actively contested within a distinctly therapeutic model of intimacy.

[1] Corresponding author: skyler.wang@mcgill.ca

## INTRODUCTION

Over the past two decades, digital technologies have come to scaffold the most intimate corners of social life. Dating platforms have restructured the social pathways through which couples meet (Rosenfeld et al., 2019), and algorithmic systems have introduced new logics of efficiency, comparability, and optimization into assortative mating (Bucher, 2018; Wang, 2025). Intimacy today is no longer simply enabled by technology; it is increasingly organized and made legible through it (Baym, 2015; van Dijck, 2013).

The recent diffusion of consumer-facing generative artificial intelligence (AI) has added a new layer of mediation. While dating apps transformed how people meet, general-purpose large language models (LLMs) like ChatGPT, Claude, and Gemini are now beginning to infiltrate everyday relationship life. Across vernacular accounts online, people describe asking these models to compose apologies, soften reprimands, mediate fights, rewrite breakup texts, and interpret partners' cryptic messages. In each scenario, AI does not substitute for the romantic partner; it slips into the space between partners, modulating how emotions are articulated.

Yet existing scholarship on AI in intimate life has principally attended to AI as a relational partner—to the companionship offered by chatbots and avatars, and to the parasocial bonds people develop with them (Pentina et al., 2023; Skjuve et al., 2021; Turkle, 2011; Wang and Dehnert, 2026). When generative AI is used not to replace a human partner but to mediate between human partners, the unit of analysis shifts from human–machine intimacy to *machine-mediated human intimacy*. While Hancock et al. (2020) outline the contours of "AI-mediated communication" (AI-MC) as an emerging field, empirical work in this area remains nascent (Hohenstein et al., 2023; Jakesch et al., 2019; Mieczkowski et al., 2021), and only recently has it turned to romantic life (see Levkovich and Alon, 2026, for a review).

This article addresses that gap and proposes a theoretical extension. Drawing on a multimodal qualitative corpus of 131 publicly available artifacts (e.g., social media posts, threads, videos, blogs, and journalism), it examines how people report, debate, and represent the use of generative AI as an emotional intermediary, and what these public accounts reveal about emotional labor, authenticity, and the structural form of contemporary romantic relationships. The conceptual work rests on two complementary contributions. The first, the *covert triad*, names one structural reconfiguration of romantic exchange, whereby what phenomenally presents itself as a relation between two is, operationally, a relation among three—with the third entity often only known to the deploying partner and not to the receiving partner. The term draws on Simmel's (1950) classical insight that the addition of a third party to a dyad does not merely add a person but transforms the relational form itself, opening it to mediation, coalition, and concealment. The covert triad is, in this sense, a contemporary Simmelian formation.

The second contribution identifies the mechanism that produces the covert triad: the delegation of a specific component of emotional labor (Hochschild, 1983) we call *articulation labor*—the expressive work of converting an inner state into language a partner can receive, as distinct from the interior work of generating the feeling itself. E. Anderson (2023) identifies a related concept, hermeneutic labor: the explicit, deliberative work of interpreting one's own and a partner's states and working through the tensions between them. In both accounts, articulation is bundled into something larger: for Hochschild, the management of feeling; for Anderson, interpretation. LLMs make articulation easier to delegate separately, so a person can retain the feeling but hand off the wording. Yet choosing those words requires anticipating how a partner will understand them, so an LLM ends up interpreting the partner too. It therefore functions less as an editing tool than as a third party that mediates the exchange. While outsourcing articulation is not new in itself (greeting cards and close friends have long supplied words for feelings), what is new here is the scale, speed, and concealability with which it can now be handed to a nonhuman, corporate-owned system.

Our argument proceeds in three moves. First, generative AI does not eliminate emotional or hermeneutic labor but unbundles them: in the recurring vernacular image of AI as "translator," the feeling is mine, the wording is co-produced. Second, articulation labor under AI mediation becomes pre-emptive and rehearsed, extending the editorial logics of text-based communication (Mieczkowski et al., 2021; Walther, 1996). Third, delegation pushes authenticity—late-modern intimacy's demand that expression be sincere and *one's own* (Illouz, 2012)—from a property of linguistic authorship toward one of emotional ownership. When undisclosed, LLM-assistance restructures the dyadic form of romantic communication into a covert triad, in which AI operates as the kind of quasi-social actor anticipated by the Computers Are Social Actors (CASA) tradition (Gambino et al., 2020; Nass and Moon, 2000; Reeves and Nass, 1996). This article contributes to scholarship on digital intimacy, emotional labor, and AI-mediated communication by showing how generative AI enters a domain long held sacred and unsettles—from the inside—what authenticity in intimate life looks like.

## MEDIATED INTIMACY AND THE ARTICULATION OF FEELING

Studies of digital intimacy have long emphasized that personal life is increasingly shaped by the computational infrastructures or platforms in which it unfolds (Baym, 2015; van Dijck, 2013). Yet most of this work, including research on dating apps and algorithmic matchmaking (Bucher, 2018; Rosenfeld et al., 2019; Wang, 2025), has focused on partner selection or search rather than on the ongoing communicative work that sustains two people after the swipe. This article extends the algorithmic-intimacy frame to a different temporal site—the daily, communicative practices through which couples make and remake their bond.

This communicative work is the object of a parallel literature on computer-mediated communication (CMC). Goffman's (1959) account of impression management—the situated, audience-aware shaping of self-presentation—has long been productive for digital scholars, since text-based communication intensifies the editorial dimensions of expression by lending it asynchrony, persistence, and revisability (boyd, 2014). Walther's (1996) hyperpersonal model formalized this insight, arguing that mediated communication often enables more strategic and idealized self-presentation than face-to-face contexts permit. From this tradition, two insights matter for our argument. First, authenticity in mediated settings is something people perform, carefully and continuously, with an audience in mind (Marwick, 2013; Marwick and boyd, 2011). This dovetails with Illouz's (2012) account of late-modern intimacy as suffused with reflexive demands, in which partners are expected to articulate their inner states consistently, to subject the relationship itself to ongoing discursive negotiation, and to do so in ways that signal authentic engagement. That demand, we stress here, is not universal. It belongs to a historically specific, class-inflected, and largely Anglophone therapeutic model of intimacy (Illouz, 2008), and our argument focuses on AI's entry into this specific cultural landscape. Second, interpersonal CMC emphasizes the labor of compensating for missing paralinguistic cues—tone, pacing, eye contact, and bodily presence must be approximated through orthography, punctuation, emoji, timing, and word choice (Hjorth and Lim, 2012; see also Gershon, 2010, on the cultural pragmatics of breakup texts). A person who rewrites a text three times before sending is engaging in the labor of articulation; the user who first runs it through ChatGPT extends this practice, with one critical difference—the editorial work is no longer fully their own.

This editorial work is itself a form of emotional labor. Hochschild's (1983) classic account—the disciplined management of feeling in line with social rules—emerged from occupational contexts but has long been extended to intimate life, where its distribution is markedly gendered (Erickson, 2005). Expression has a place in her account, but a narrow one: her distinction between surface and deep acting separates display from feeling only where the two diverge; putting a genuine feeling into words falls outside that distinction. E. Anderson (2023) comes closer. Her concept of hermeneutic labor is reflective rather than real-time work: it includes understanding one's own feelings and "presenting them in an intelligible fashion to others" (2023: 178), reading a partner's, and weighing the two to resolve conflict. But for Anderson, expressing feelings is one part of interpretation, not work in its own right. We therefore distinguish *feeling labor* (the interior work of generating and inhabiting emotion in line with relational expectation) from *articulation labor* (the expressive work of converting that felt experience into language a partner receives).

Of the two, only articulation is readily transferable. No one can feel on another's behalf, but a message is detachable: someone else can help draft it, and the sender can send it as their own. Face to face, feeling and articulation are harder to separate, but asynchronous, revisable text has long let them drift apart. The gap such delegation fills is long-standing and gendered. E. Anderson notes that men are often socialized not to put feelings into words ("normative male alexithymia"; 2023: 186), while women turn to friends to decode men's "scarce words" and craft replies, in what she calls "hermeneutic labor workshops" (2023: 190–191). LLMs make the drift between feeling and articulation ever more visible by folding the draft and the friend into a single tool, one available to people who were never taught to put their feelings into words. Articulation labor, already the most detachable component of emotional labor, is likely the one generative AI reorganizes most heavily.

## THE COVERT TRIAD: AI AS INTERMEDIARY IN INTIMATE LIFE

A small but growing literature has begun to address the intersection of AI and intimate life. The most developed strand concerns AI as a relational partner—chatbot companions such as Replika, Character.AI, and Xiaoice that occupy quasi-romantic roles in users' lives (Pentina et al., 2023; Skjuve et al., 2021; Turkle, 2011), or what Wang and Dehnert (2026) term the "on-demand intimacy" of sociotechnically engineered systems that provide around-the-clock affection. Hancock et al. (2020) propose a distinct analytic frame—AI-mediated communication (AI-MC)—to describe communication between humans in which an AI agent intervenes to shape or generate a message. Empirical work in this tradition has examined how AI assistance alters trust judgments, interpersonal language and relational perception, and assessments of sincerity (Hohenstein et al., 2023; Jakesch et al., 2019; Mieczkowski et al., 2021). Cross-cutting both literatures, the Computers Are Social Actors (CASA) paradigm (Gambino et al., 2020; Nass and Moon, 2000; Reeves and Nass, 1996) holds that people apply social scripts to computational systems even when they are fully aware they are interacting with a machine. We situate this article at the intersection of AI-MC and intimate-life scholarship, but argue that something more than an "intervention" or "assist" is at stake. What the AI-mediated romantic exchange produces is a structural transformation of the relational form itself, for which the most useful vocabulary is Simmel's sociology of the dyad and the triad.

For Simmel (1950), the dyad is the most distinctive of all social forms and, arguably, the most fragile. It admits no third party, mediator, or audience. Each partner faces the other directly; the relation has no existence independent of the two who sustain it, and it dissolves the moment one withdraws. The dyad is also, for that reason, the form most heavily invested with intimacy in modern Western thought—its very fragility is what makes it precious. Romance, in particular, has long been theorized as the dyadic relation par excellence—two individuals constituting a "world of two," irreducible to anything outside themselves (Berger and Kellner, 1964; Luhmann, 1986).

What Simmel saw was that the triad makes structurally available what the dyad forecloses—mediation, coalition, and the use of the third as an arbiter. The third party can play several roles—the impartial mediator who reconciles a quarrel by "depriv[ing] conflicting claims of their affective qualities" and "neutrally formulat[ing] and present[ing] these claims to the two parties involved" (Simmel, 1950: 147), the *tertius gaudens* who profits from the division between the other two, or the divide-and-conquer figure who actively engineers their separation. Each role transforms the dyadic relation differently, but all share a structural feature—once a third is introduced, the relation between the original two is no longer self-contained. It is now visible to and partially constituted through someone outside it. Importantly, Simmel also recognized that the triad opens vulnerabilities the dyad does not—it permits coalition formation (two against one) and strategic concealment (one withholding from another what is shared with the third). We extend this insight one step further to the possibility that what passes between two members may be shaped, without their full awareness, by a third who is not co-present but concealed.

AI-mediated romantic exchange, this article argues, is a Simmelian triad of a distinctive kind. It is covert, in the sense that the third party's participation is typically known to only one of the two human partners; it is asymmetric, as the deploying partner experiences a triadic configuration while the receiving partner experiences a dyadic one; and it is machinic, in that the third is not a human ally with motives of their own but a model with no relational stake. A couple's therapist is a visible, mutually acknowledged third, while a friend in whom one partner confides is a relational third whose existence each partner can interrogate. The AI intermediary is neither. Its participation in the production of an utterance is invisible by default and visible only by disclosure—and, as the findings below show, disclosure is often elided. This is precisely the condition Simmel made constitutive of intimacy—its "peculiar color" exists, he argues, only where "the whole affective structure is based on what each of the two participants gives or shows only to the one other person and to nobody else" (Simmel, 1950: 126). The covert third quietly violates that condition while leaving the dyad's surface form undisturbed.

Two qualifications bound this claim. First, not every use of AI makes a triad. When AI only fixes spelling or suggests synonyms, it remains an instrument rather than a participant, and the dyad stays intact. AI becomes a third when it takes up interpretive work—reading the partner, weighing the situation, proposing what should be said and why—because it then brings a perspective of its own to bear on the relationship. Second, covertness is a matter of degree. We focus on the covert configuration not because it is the only formation but because its asymmetry most directly unsettles the intimate dyad, and we return to other forms in the discussion.

## METHODS

This study uses qualitative thematic analysis to examine how the use of generative AI as an intermediary in contemporary romantic communication is publicly reported, debated, and represented. The empirical material comprises 131 artifacts drawn from online forums, social media, and long-form blogs or journalistic accounts—a multimodal corpus of publicly available content produced between 2023 and 2026, the period spanning the rapid mainstreaming of consumer-facing LLMs. Because the study's aim is to build theory, the corpus supports analytic generalization—developing concepts that explain how AI mediation works in romantic communication—rather than claims about representativeness or prevalence (Yin, 2018). Sampling was purposive, favoring varied, information-rich cases of AI-mediated practice. Given that our search terms were oriented toward difficulty and discovery (e.g., "boyfriend using AI to text me"), the corpus likely over-represents covert and contested use. Since open and mutual use also appear, we treat the covert triad as one configuration among several.

Forum material constitutes the base of the corpus. The relative anonymity of platforms such as Reddit supports candid disclosure of intimate experiences that would be difficult to elicit through interviews or surveys (Andalibi, 2020). Posts were drawn from subreddits with active discussions of relationships, communication, and AI use—including r/relationships, r/relationship_advice, r/dating, r/AmIOverreacting, r/amiwrong, r/OpenAI, r/ChatGPT, and r/therapyGPT—using search terms such as "used ChatGPT for apology," "AI breakup message," and "ChatGPT helped me text." Of approximately 200 candidate Reddit posts initially identified, 55 met the inclusion criterion of explicitly describing the use of a generative AI tool to draft, edit, or interpret a message in a human-to-human romantic context. Threads concerning AI companions as partners were excluded.

Beyond public forums, the corpus included video and long-form written material. Video sources spanned long-form YouTube content (video essays, reaction videos, satire, and promotional videos) and short-form material from TikTok and Instagram Reels, identified using search terms such as "ChatGPT relationships," "AI wrote my apology," "boyfriend using AI to text me," and "ChatGPT in a fight with my girlfriend." Of the 300 candidate videos, 41 met the inclusion criterion. Video is analytically useful because creators routinely demonstrate AI use on-screen—pasting partner messages into chatbots, reading aloud AI-drafted replies—making user-side reasoning visible in ways that flat text cannot. Long-form written material (n = 35) includes first-person essays on Medium and feature reporting in venues such as *The Guardian*, *Vice*, *Business Insider*, *and The Chicago Tribune*. These pieces let us examine how the cultural meanings of AI-mediated intimacy are publicly negotiated.

Because these sources differ in what they can show, we coded each artifact by data type as well as by theme, distinguishing four kinds of evidence that cut across platforms (Table 1; see Butkowski and Corry, 2025). More specifically, first-person user accounts show practice as users report and justify it, while receiving-partner accounts show how AI use is discovered and received. Commentary, journalism, and advice show how others talk about and judge AI use, rather than how people use it. Satire and promotional content, such as South Park and a Hapday video, are cited as examples of how the culture imagines AI in relationships. The full corpus was analyzed using a thematic, inductive coding procedure (Braun and Clarke, 2006). An initial close reading of all material generated open codes—including tone management, message rehearsal, clarification of feeling, interpretation of partner messages, concerns about deceit, and justification of use. Codes were iteratively refined and clustered into three overarching themes that organize the findings section. To enhance analytical rigor, disconfirming evidence was actively sought (Tracy, 2010); we report the principal cases—for example, open or mutual use—alongside the themes they qualify.

**Table 1.** Corpus composition by data type and theme (number of artifacts).

| Data type | Treated as evidence of | All | Theme 1 | Theme 2 | Theme 3 |
|---|---|---|---|---|---|
| First-person user accounts | Practice as reported and justified | 62 | 41 | 37 | 22 |
| Receiving-partner accounts | Discovery and reception | 26 | 3 | 5 | 24 |
| Commentary, journalism, and advice | Public framing and normative judgment | 36 | 18 | 12 | 27 |
| Satire and promotional content | Cultural imaginary | 7 | 4 | 3 | 2 |
| Total | | 131 | 66 | 57 | 75 |

*Note.* Artifacts may anchor more than one theme. Theme 1 = motivations; Theme 2 = rehearsal and interpretation; Theme 3 = authenticity and the covert triad.

Following established guidelines for research with online communities (franzke et al., 2020), Reddit usernames have been removed, and quotations have been lightly paraphrased to reduce traceability. Video creators and journalists, who publish under their own names, are quoted and attributed following standard academic practice. The analysis treats user accounts as situated narratives rather than transparent reports of inner experience or behavior.

## FINDINGS

As noted above, we develop three themes that map how generative AI is being absorbed into everyday romantic communication. The first considers the motivations behind the delegation of articulation labor—what users describe as the 'problem' that AI is called in to solve. The second concerns the procedural form articulation labor takes under AI mediation—how, concretely, users do it, and how the practice extends and transforms the editorial logics of earlier CMC. The third turns to consequences, first for authenticity, which becomes a contested and shifting boundary, and then for the structure of romantic exchange. When AI use goes undisclosed, that structure becomes a covert triad.

**Outsourcing Articulation: Motivations and the Unbundling of Emotional Labor**

The most pervasive way people in our corpus framed AI use was as *translation*—a means of converting inner emotional states into externally communicable language, and thus precisely the work of articulation labor. One Reddit user offered a characteristic formulation: "I knew I was sorry; I just couldn't find a way to say it that didn't come off as defensive" (Reddit, r/dating, 2023). Another wrote, "The feeling was already there—I just needed help putting it into words that wouldn't make things worse" (Reddit, r/relationship_advice, 2024). In both cases, the difficulty named is not emotional absence but the gap between one's feelings and the language needed to express them to a partner.

This same logic recurs across our corpus. In a widely shared Medium essay, the writer Stephen Anderson describes a sequence in which he asks ChatGPT to draft an apology to his wife and later to compose a "heartfelt reply" to her text on his behalf. The essay closes, half-joking and half-not, with "I love you too, ChatGPT" (S. P. Anderson, 2023). The translational frame is even more explicit in a promotional YouTube short that advises viewers who have been broken up with to "try AI to help with your relationships," demonstrating the workflow of pasting a partner's message into a chatbot and asking it to draft a response (Hapday, 2024). In a *Guardian* feature on the rise of AI-assisted personal messaging, the reporter observes that "people are outsourcing emotionally difficult conversations to AI tools" and that the tool is "becoming the invisible infrastructure of personal communications… even though we associate such compositions with 'from the heart' authenticity" (Trapani, 2025). Across data types, the recurring trope is one of inarticulacy redeemed—the speaker has feeling but lacks fluency. In other words, the machine supplies form, not content.

A second, less common motivation begins earlier in the process. Rather than seeking words for a feeling they already have, some users turn to AI to work out what the feeling is. One Reddit user writes: "I couldn't tell whether I was angry or just hurt until I started typing it out to ChatGPT and it asked me something back" (Reddit, r/relationship_advice, 2025). Users describe the model as giving an inchoate feeling a provisional shape—a candidate sentence they can accept, revise, or reject—and they claim the feeling that crystallizes around it as their own. These accounts also show that the components of emotional labor are not sealed off from one another. The words that help a user recognize a feeling can also color it, a possibility participants in other research raise directly when they worry that AI can "put words in users' mouths, and convince them they feel more or less than they really do" (Tseng and Liang, 2026: 10). Articulation, in other words, does not only carry feeling; at the edges, it helps form it.

The translational frame allows users to retain emotional ownership while delegating linguistic precision. Sociologically, this is significant: the accounts examined here bring to the surface a division that Hochschild's (1983) account of emotional labor leaves implicit. The interior side—feeling remorse, vulnerability, care—remains the user's. The exterior face—converting that feeling into a credible utterance—is co-authored by AI. The distinction between feeling labor and articulation labor proposed earlier is, in this sense, not only an analytic refinement but a distinction users themselves draw when they account for their practice. The labor of intimate communication is thus reorganized as a hybrid accomplishment—human in its interior form, sociotechnical in its outward expression.

**How Articulation Labor Is Done: Rehearsal and Interpretation**

While the first set of findings highlights why articulation labor is delegated, the second concerns how it is performed once delegated. Our data suggest AI mediation makes articulation labor anticipatory and rehearsed. More specifically, users on Reddit describe turning to AI not only to phrase but to simulate and test their communication before sending it off. One user explained: "I didn't want to make the fight worse, so I ran what I'd written through ChatGPT first" (Reddit, r/ChatGPT, 2025). Another wrote: "I kept checking with it whether my message came across as too cold or too harsh before sending. I don't want to seem insensitive" (Reddit, r/dating, 2026). The orientation here is pre-emptive—not just did I say what I meant, but will my partner read this the way I intend?

The pattern is familiar enough to be satirized. The South Park episode "Deep Learning" (with clips viewed by millions on YouTube) parodies the process—a character repeatedly regenerates ChatGPT's draft, accompanied by the now-vernacular instruction to "make this sound less aggressive" (Parker and Stone, 2023). Several YouTube creators treat the practice as a skill to optimize: one dating coach, for example, runs the same dating-app exchanges through three chatbots and concludes that ChatGPT writes the best messages, including some he "personally would copy and paste and send" (High Integrity Skills, 2025). Journalistic coverage echoes the pattern, with reporting on users who describe AI as "a buffer," a "sanity check," or a "first reader" before sending difficult messages (Trapani, 2025).

What emerges is a model of articulation labor that is pre-tested rather than spontaneous. LLMs offer a rehearsal function, a sandbox in which users iterate with a simulated partner to manage the affective valence of the actual exchange. This pattern resonates with arguments in the AI-MC literature about the extension of selective self-presentation (Hancock et al., 2020; Mieczkowski et al., 2021), but it also points to something more specific. The labor of articulation, once carried out privately in the head or with the help of an accessible friend, now unfolds in dialogue with a system that can generate responses, anticipate how they might be received, and offer alternatives at the speed of conversation.

The ability to rehearse operates in two directions. It is used not only to prepare one's own messages but also to interpret the partner's. A recurring scenario in our data involves users routing an ambiguous text from their partner through AI for interpretive assistance. One Reddit user writes: "I asked ChatGPT what my partner probably meant by the text because I couldn't work out what he was saying—is he upset with me?" (Reddit, r/ChatGPT, 2023). Another reports: "It honestly helped me settle down before I answered. Kind of like talking it through with somebody" (Reddit, r/relationship_advice, 2025). In a first-person *Business Insider* essay, the writer describes an extension of this use—uploading the full chat history of her relationship and asking the model to evaluate the couple's communicative dynamics, only to be told that her boyfriend was "a better communicator" than she was (Liu, 2025). Articulation labor, in other words, begins before a word is written: to reply well, the user first has to understand what they are replying to. This is where articulation shades into what E. Anderson (2023) calls hermeneutic labor—discerning a partner's feelings and intentions and working out how to respond. What the user produces with AI is not yet a reply but a reading—a working account of what the partner meant that they weigh against their own before responding. It is also where consent is most at stake. The partner's words are shared with a third party they did not choose, a problem some users sense themselves, as when an interviewee in Fu et al.'s (2025) study balked at pasting a partner's messages into AI because "it's also their texts" (2025: 20).

The rehearsal function also tends to standardize. Users repeatedly described AI outputs as smoother, more therapeutic, and more "I-statement"-inflected than their own first drafts. As one Reddit user puts it, "all those long, emotional messages started to sound identical" (Reddit, r/relationship_advice, 2025), describing this homogenization from the receiving partner's vantage point. A news segment on AI in romantic communication captures the same intuition, noting that "people are now editing emotions like drafts" (CBS News, 2024).

This pattern also sits uneasily with normative ideals of spontaneity that suffuse romantic discourse. As late-modern intimacy demands both reflexivity and authenticity (Illouz, 2012), the easy access to rehearsing with an artificial agent intensifies the former while complicating the latter. Users themselves underscore this tension—many of the accounts examined here oscillate between describing AI as a prudent safeguard and worrying that pre-tested speech is, by definition, less honest.

**Consequences: Authenticity Reconfigured, the Dyad Restructured**

The delegation of articulation labor has two related consequences that our data point to: a normative reconfiguration of authenticity and a structural reconfiguration of the romantic dyad into the covert triad. Both follow from the same underlying shift. Once articulation labor can be delegated, the form of an utterance no longer carries a reliable trace of its author; once a concealed third party participates in the production of intimate speech, the conversational dynamic is no longer dyadic in the way it appears.

Across the corpus, authenticity emerges as a contested site. Many users insist on a distinction between authenticity of feeling and authenticity of phrasing, locating sincerity in the former. "The feelings are mine," one user writes, "the wording just helps get across what I'm trying to feel" (Reddit, r/relationship_advice, 2023). Another argues: "It's still real if I mean what I'm saying, even when something helped me word it better. I didn't want to come across as mean" (Reddit, r/ChatGPT, 2024). A commenter defending the practice on Reddit compares it to "picking out a nice greeting card—it's the thought behind it that counts" (Reddit, r/relationship_advice, 2024).

Some, however, reject this division between emotional ownership and linguistic authorship and treat AI use as inauthentic by definition. "If you need AI to write your apology, then it isn't really coming from you," one user contends (Reddit, r/therapyGPT, 2024). Here, authenticity is reattached to the labor of expression itself. The repudiation is often sharpest when the receiving partner discovers the delegation. A widely engaged thread describes a wife learning that her husband had used ChatGPT to write "every card he's given me (Valentine's, Mother's Day, and so on) and for every apology he's made to me over the past several months." The top comment of the thread frames the issue as deception more than as mediation: "It's the lying about it that would be the dealbreaker for me. So he used the plagiarism machine to churn out fake love notes and let you believe the words were his own" (Reddit, r/relationship_advice, 2024). Another receiving partner asks plaintively: "Couldn't he have put even a LITTLE soul into the apology? I'm not buying some fake apology from a robot" (Reddit, r/relationship_advice, 2026). A third writes: "I feel tricked and betrayed. I opened up and was vulnerable with him about my feelings, and his responses were AI-generated" (Reddit, r/AmIOverreacting, 2023). In these cases, the discovery of AI mediation does not merely surprise the receiving partner—it retroactively reframes the entire history of the exchange as inauthentic. Press coverage amplifies this framing; a *Guardian* feature on AI use in dating-app conversations describes users who felt "deceived" after learning intimacy had been AI-assisted, a phenomenon the reporter terms "chatfishing" (Hood, 2025).

The fault line between these positions traces an older debate in CMC scholarship over whether mediated communication is essentially performative or expressive (Marwick, 2013; Marwick and boyd, 2011). What was once a theoretical debate is now being waged in real time by ordinary users themselves. Authenticity is not a stable property of communication but a normative boundary actively being redrawn. The boundary is contested at multiple sites—between feeling and form, between effort and outcome, and between disclosure and concealment. As one advice columnist puts it, surveying the dilemma posed by a reader whose boyfriend has used ChatGPT to write romantic texts: "How much of him is actually in these messages? Is AI polishing his communication, or generating it?"—a question, she adds, sharpened by the fact that "our very notions of 'reality' are shifting by the minute and we can no longer tell immediately what's real and what's not" (Pulley, 2026).

Discovery, however, is not the only way AI use comes to light, and receiving partners do not uniformly object. Although many users in the corpus describe not telling their partner that AI helped draft a message, those who do disclose report mixed reactions—some partners are reassured by the user's effort to "get it right," while others read the disclosure as a betrayal. In one notable case, a woman wrote that although she was annoyed her husband had needed ChatGPT to write to her, it "created an opportunity to open a door for some tough conversations," and she ultimately took it as a sign that "he cared" (Reddit, r/AmIOverreacting, 2025). Reactions like hers are the exception, however; negative responses to disclosure far outnumber positive ones in our corpus. What separates the two appears to be less the AI use itself than what the partner takes it to mean: evidence of care and effort, or a sign that the sender handed off something that should have been theirs to say. At the other extreme are couples for whom AI use was never hidden at all. One user described how she and her boyfriend, stuck in a recurring argument about chores, turned on ChatGPT Voice mode and asked it to "translate" each side to the other. "It was weirdly helpful," she wrote, "because neither of us had to be the bad guy" (Reddit, r/relationships, 2026). Here the AI is a visible, shared third, closer to a couple's therapist than a ghostwriter, and the question of disclosure never arises.

The delegation of articulation labor can also reshape the structural form of romantic communication. Users begin by describing AI as a tool for drafting—many go on to describe it as a confidant, interpreter, sounding board, or even a friend. The interpretive pattern documented above is where this drift becomes visible in practice: AI shifts from instrument to *interlocutor*. When the user quoted earlier describes AI as "kind of like talking it through with somebody, " the word "somebody" is telling. AI is treated as a partner in the conversation, not a tool for it.

Receiving partners register the shift most sharply. A *Vice* piece describes a man whose girlfriend "won't stop using ChatGPT for relationship advice," and he reports feeling "ambushed with the thoughts and opinions of a robot" during their arguments (Lavin, 2025). The partner has, in effect, brought a third interlocutor into every disagreement. A Reddit user makes the structural complaint explicit: "Every time we argue, the first thing he does is ask ChatGPT… ChatGPT doesn't have feelings, I DO" (Reddit, r/amiwrong, 2025). A widely circulated YouTube video opens with the line: "It felt like he optimized the text instead of actually talking to me" (Stronge, 2024).

This quasi-social positioning is anticipated by the CASA paradigm (Nass and Moon, 2000; Reeves and Nass, 1996), recently updated by Gambino et al. (2020) to account for advances in conversational AI. The CASA thesis contends that people mindlessly apply social scripts to computational systems that exhibit social cues, even though they know those systems are not human. What users in the present corpus exhibit is consistent with that prediction, but with a twist. The social scripts being applied are not generic—they are specifically those of intimate confiding and therapeutic listening. AI is treated less as a system one operates than as a friend one consults and less as a search engine one queries than as a confidant one updates.

Where AI use goes undisclosed, what appears to be a relationship between two people becomes a covert triad. AI does not enter as a relational competitor (it is not, in these cases, a chatbot companion) but as a relational scaffold. The user processes the partner's behavior, drafts responses, calibrates affect, and in some accounts even learns about themselves—all through AI. The triad forms not because AI replaces the partner, but because it shapes what the user says, with only one partner aware of its role.

## DISCUSSION AND CONCLUSION

Drawing on public accounts, commentary, and representations of AI-assisted romance from 2023 to 2026, this article makes three arguments. First, people turn to AI chiefly because converting felt experience into language a partner can receive is hard. Second, under AI mediation, articulation labor becomes pre-emptive, rehearsed, and standardized. Third, this delegation moves authenticity, for users if not always for their partners, from linguistic authorship toward emotional ownership and, where undisclosed, turns the romantic dyad into a covert triad.

Distinguishing feeling labor from articulation labor matters for two reasons. For one, it specifies what AI is changing. Rather than asking whether AI "replaces" emotional labor—a question to which the answer is no (or not yet)—the framework asks which component is delegated, under what conditions, and with what consequences. Furthermore, although our interpretive design cannot test them, the distinction yields propositions that future work could. If it holds, AI assistance should track variation in articulation difficulty more closely than variation in felt intensity: apologies, conflict de-escalation, and ambiguous reproach (i.e., high-articulation-cost scenarios) should invite AI assistance more than emotionally intense but linguistically formulaic contexts.

The boundary cases above show where the distinction could break down. When a user discovers through typing to ChatGPT that they are “hurt” rather than “angry,” the words shape the feeling, not just express it. If that happens often, feeling and articulation are better seen as a spectrum than as two separate kinds of work.

The findings also indicate that many users now locate authenticity in emotional ownership rather than linguistic authorship. In romance, however, the form of an utterance carries unusually heavy normative weight (Luhmann, 1986). An apology or a declaration of love does not simply report a feeling; saying it is part of the act. Outsourcing the saying can therefore look to the partner on the receiving end like altering the act itself. The articulation-labor framework helps explain why this reconfiguration is contested. Users who anchor authenticity in emotional ownership treat articulation labor as legitimately delegable, whereas users who anchor it in expressive effort treat the delegation as a forfeiture. Both positions are coherent within the framework, and the disagreement among users reflects a live cultural negotiation. Importantly, this negotiation is not symmetrical. The party deploying AI knows it has been used; the receiving party often does not. Authenticity, in this asymmetric configuration, is bound up with disclosure. That said, disclosure is not a simple fix. Experimental work on AI-assisted intimate messaging finds that greater AI involvement weakens receivers’ sense of a message’s effort, authorship, and authenticity, and that brief sender disclosure only partly offsets that loss (Fan et al., 2026). The imbalanced structure of AI-mediated romantic exchange, thus, strains conventional norms of transparency in intimate life.

Structurally, the findings show how romantic communication—conventionally theorized as dyadic—can take the form of a covert triad. Read through Simmel (1950), this is not an additive change but a transformation of relational form. The covert triad makes mediation, coalition, and concealment available at once, with AI mediating between partners, aligning in the deploying partner’s experience against the difficulty of saying hard things, and remaining concealed from the receiving partner. Yet covertness is one position on a continuum of visibility. Our corpus contains examples of open use, where partners are made aware of LLM use in everyday romantic communication. Existing studies describe partners imagining a consent-based AI mediator present to both, closer to Simmel’s impartial arbiter (Fu et al., 2025), and, at the other extreme, couples who each consult their own model until one partner’s “my AI says you’ve been a bad boyfriend” turns the arrangement adversarial (Tseng and Liang, 2026: 11). As disclosure spreads, the triad need not collapse back into a dyad; it can open onto the other configurations Simmel anticipated, from open mediation to rival coalitions.

Taken together, these findings show how AI mediation can reshape a culturally specific model of intimacy. In the public accounts examined here, intimacy becomes a sociotechnical accomplishment: produced in concert with—and partly through—the tools that mediate its expression. The homogenizing pull of LLM outputs toward smoother, more therapeutic, more "I-statement"-inflected speech is more than stylistic sameness. Read through Illouz's (2008) account of the therapeutic ethos, it is the mass diffusion of one culturally particular—therapeutic, largely Anglophone, middle-class—model of emotional expression, the model that advice columns and self-help literature have long promoted (Cancian and Gordon, 1988; Illouz, 2008) and that a handful of frontier models now supply on demand. The standardization of intimate language is thus a question of political economy as well as culture. Even users seeking advice rather than drafts notice this sameness; participants in Brailas and Tsolakis's (2025) study observed that ChatGPT's relationship guidance followed a recognizable template. If millions of apologies are routed through a small number of base models, what becomes of idiosyncrasy and voice? Much of what gives intimate language its force is precisely that it is risky and imperfect—that it bears the marks of a particular speaker's struggle to articulate what they mean.

Beyond romance, the distinction suggests a broader question about contexts in which expression can be detached from its source. Students drafting essays and professors writing recommendation letters with an LLM may face a comparable ambivalence: the ideas are theirs, the prose is not—and they remain uncertain about whether the result is truly their own. What we unveil is the mechanism behind that intuition. Just as romantic users decouple the feeling they own from the wording they delegate, professionals and knowledge workers may decouple intellectual labor—the work of having an argument or insight—from articulation labor, the work of rendering it in fluent, situationally appropriate language. In each domain, authenticity may be located in ownership of the ideas rather than in authorship of the words.

This study is not without limitations. Our corpus is heavily Western-centric and skews toward platforms whose user bases tilt young, comparatively educated, and digitally fluent; how AI mediation is taken up where authenticity is not anchored in self-authored expression is an open comparative question. Because current AI systems are built for individual users, the evidence is weighted toward the partner who consults the system rather than the one who receives its output (Levkovich and Alon, 2026). The gendered dimension of the phenomenon also could not be addressed here, though E. Anderson's (2023) work suggests it is central: delegating articulation and hermeneutic labor may redistribute a burden women have disproportionately carried or add a new layer to it. Future research should involve perspectives from both partners.

Finally, the covert triad should not be read as a stable end state. One potential outcome of all of this is normalization: if most intimate messages come to be assumed to pass through an LLM, the triad would cease to be covert. But users also domesticate, repurpose, and resist technologies (Siles, 2023). In other words, the instrumentality and sameness that trouble users here could turn unassisted, imperfect, and even "painful" messages (Tseng and Liang, 2026: 10) into new markers of authenticity—a countermovement our corpus does not yet show. Whether authorship stops mattering or comes to matter for new reasons—less who wrote the words than what it cost to find them—is a question these early accounts can raise but not yet answer.